# Disclosure Control of Machine Learning Models from Trusted Research Environments (TRE): New Challenges and Opportunities

Emily Jefferson(1),Esma Mansouri-Benssassi(1)*, Simon Rogers(2)*, Jim Smith(3), Felix, Ritchie(3),

Maeve Malone (1),

(*) These authors contributed equally to this work

(1) University of Dundee

(2) NHS National Services Scotland

(3) University of the West of England

**Dr Esma Mansouri-Benssassi** : Health and Clinical Services, School of Medicine, University of Dundee , Dundee, Scotland.

**Dr Simon Rogers** Artificial Intelligence Centre of Excellence, NHS National Services Scotland , Glasgow, Scotland.

**Prof Jim Smith**, FET - Computer Science and Creative Technologies, University of West England, Bristol, England.

**Prof Felix Ritchie** FBL - Accounting, Economics and Finance, University of West England, Bristol, England.

**Maeve Malone**, School of Social Sciences, University of Dundee, Dundee, Scotland.

**Prof Emily Jefferson**: Population Health and Genomics, School of Medicine, University of Dundee, Dundee, Scotland.

Keywords: Trusted Research Environment, AI, Machine Learning, Data Privacy

Word count: 5299

# Abstract

## Objective

Artificial intelligence (AI) applications in healthcare and medicine have increased in recent years. To enable access to personal data, Trusted Research environments (TREs) provide safe and secure environments in which researchers can access sensitive personal data and develop Artificial Intelligence (AI) and Machine Learning models. However currently few TREs support the use of automated AI-based modelling using Machine Learning.

Early attempts have been made in the literature to present and introduce privacy preserving machine learning from the design point of view [1]. However, there exists a gap in the practical decision-making guidance for TREs in handling models disclosure.

Specifically, the use of machine learning creates a need to disclose new types of outputs from TREs, such as trained machine learning models. Although TREs have clear policies for the disclosure of statistical outputs, the extent to which trained models can leak personal training data once released is not well understood and guidelines do not exist within TREs for the safe disclosure of these models.

## Materials and Methods

In this paper we introduce the challenge of disclosing trained machine learning models from TREs. We first give an overview of machine learning models in general and describe some of their applications in healthcare and medicine. We define the main vulnerabilities of trained machine learning models in general. We also describe the main factors affecting the vulnerabilities of disclosing machine learning models. This paper also provides insights and analyses methods that could be introduced within TREs to mitigate the risk of privacy breaches when disclosing trained models.

## Results

Although specific guidelines and policies exist for statistical disclosure controls in TREs, they do not satisfactorily address these new types of output requests; that is, trained machine learning models. We present an analysis of the main challenges faced when disclosing machine learning models from TREs. We also propose novel directions that can be applied by TREs for risk mitigation in machine learning models disclosure.

## Conclusions

ML models do not fit existing models of output disclosure protection; there is an urgent need to address this to ensure AI can be used effectively. There is significant potential for new interdisciplinary research opportunities in developing and adapting policies and tools for safely disclosing machine learning outputs from TREs.

# Introduction

In many countries, access to sensitive data such as personal such as financial or individuals' health records, is provided to researchers and other developers through Trusted Research Environments (TREs) [2]. TREs provide a secure computational platform, within which researchers can access and analyse data, before requesting the disclosure of any aggregate level, anonymised results (for, e.g. publication).

Output disclosure processes were created with the disclosure of traditional statistical outputs in mind (for example, graphs and tables), that can be read and understood by human output checkers. The increasing popularity of Artificial Intelligence, and in particular Machine Learning methods, throws up significant challenges to these processes.

Machine Learning (ML) is a subfield of Artificial Intelligence in which algorithms are trained to perform specific tasks by exposing them to large quantities of relevant data. It is widely considered that ML systems will play an increasing role across society. It comes as no surprise then that there is an increased interest in training ML models on the data that is held within TREs. This, in itself, does not represent a substantial challenge to TREs, but in order for the trained model to be used it must be removed from the TRE and current disclosure processes are unsuitable for this type of output.

In this opinion paper, we set the scene for future work in this area. We start by describing TREs, and then introduce some popular ML models and what kind of information is present in a saved ML model. We then highlight the ways in which a trained ML model can disclose potentially sensitive information regarding the data on which it is trained and describe why this represents a disclosure risk to TREs. Finally, we summarise the areas in which we feel research is needed to provide TREs with the knowledge and tools required to safely disclose trained ML models.

# Trusted Research Environments

Trusted Research environments (TRE)s are secure and safe platforms that enable researchers to access and analyse personal data, [2]. An example within the UK is the network of Scottish research Safe Havens, in which researchers can access a wide range of health records for Scottish individuals [3]. TREs must follow specific principles when providing researchers with access to data, defined around five distinct areas as outlined by [4]. These are:

- Safe people: Individuals allowed to access data through the TRE must sign a term of use committing them to follow certain guidelines and rules. Users agree not to re-identify individuals and not to share their TRE access with non-authorised persons. They also commit to reporting any safety or security weakness identified within TREs.

- Safe projects: TRE operators need to ensure the projects' use of data is appropriate and in line with public benefits.

- Safe settings: sensitive data is held only within a secure environment, accessed through restricted areas.

- Safe outputs: TREs must set up and implement process to only allow appropriate output data to be released.

- Safe data: In general, TREs operate with highly sensitive/identifiable data: the other elements (people, projects, settings and outputs) provide a high degree of protection so that even the most sensitive data can be analysed lawfully and ethically [5].

TRE output guidelines are typically aimed at traditional statistical outputs such as aggregated results, graphs, and tables. Such outputs are human readable and can therefore be manually assessed by TRE output checkers to ensure that they do not disclose any sensitive information.  Yet, TREs have typically not had to deal with outputs that are not human readable, and not amenable to manual validation. This then risks removing or weakening a third pillar of the 5-safes - "safe outputs", unless new assistance can be provided to TRE output checkers.

Although a researcher training a ML model within a TRE may wish to remove traditional results of their analysis from the TRE, they are also likely to be interested in removing the trained ML model itself, especially if their goal is to deploy this model into a usable pipeline. A trained ML model is typically not stored in a human-readable format and, even if it were, the vast majority of models are not amenable to manual checking – they are simply too complex.  Determining how to categorize these models (in terms of AI systems), the levels of risk involved (including security risk), their transparency and explainability, the checks and safeguards needed (e.g. human and public oversight), are important initial considerations to make [6]. To understand better how this issue is being addressed in practice, in a recent survey, we interviewed 14 UK and 6 international TREs, [7] to discover current processes within TREs for AI algorithms disclosure. Our findings showed that across the board, TREs did not have mature processes, tools, or an understanding of disclosure control for AI algorithms, relying on a degree of manual checking that is likely not fit for purpose. In addition, the disclosure risks posed by trained models can be subtle, and just because an output checker cannot "see" something disclosive within the model, does not remotely suggest that it is safe.  This demonstrates a worrying gap between how ML techniques are being developed and the governance and oversight of the TREs is conducted. The gap can result in higher risks to privacy, data protection and good governance of TREs.

We will describe in more detail the unique disclosure risks posed by trained models, but first provide an introduction to some common Machine Learning models, and their application to medical and healthcare data.

## Machine Learning in Healthcare

The combination of algorithmic advances, and the availability of large datasets has led to an increased interest in the use of AI, and in particular ML. This interest is particularly high in the medical and healthcare fields where the potential for the adoption of AI and ML approaches within healthcare systems has high potential to improve efficiency, reduce the burden on clinicians, and assist in the movement to a more predictive and preventative philosophy to care [8]. However, this interest and use of AI must be tempered by due diligence of AIs strengths and limitations [9].

ML is a subset of AI, that encompasses algorithms that automatically *learn* patterns from datasets. It can be used to help humans better understand complex data, or make predictions based upon new, unseen data. ML is now a mature field, but in recent years particular technological advances, coupled with the increasing availability of very large datasets has resulted in a surge in popularity. Modern ML encompasses many types of learning and computational algorithms, [10].

ML models can be split into various categories, based on the type of problem they are attempting to solve, summarized below:

**Supervised Learning:** When performing supervised learning, algorithms are supplied with a corpus of training examples, each consisting of some input values associated with an output category or value (a *label*). The goal is to learn a mapping between the input and output. This type of learning is one of the most common and has applications such as classification and diagnosis of medical conditions, [11] and predicting risks of future health events. The output is typically either one of a number of distinct categories (e.g., case vs controls; known as classification), or a real value (regression). Supervised learning has been used with a wide range of healthcare and medical data, such as medical images, [12], Electronic Health Records (EHR), [13] and genomic data, [14] .

**Unsupervised Learning:** In unsupervised learning, the model uses unlabeled data to discover patterns in a dataset. One of the most common examples of unsupervised learning is clustering, [15]. The goal of clustering is to group data instances into clusters such that the instances sharing a cluster are similar to one another.  Unsupervised learning techniques such as clustering are often used for exploratory data analysis in healthcare and medical applications. For example, [16] used k-means clustering for identifying several subtypes of Alzheimer's using Electronic Health Records (EHR).  More recently unsupervised learning was used to learn appropriate features for COVID-19 diagnosis from CT medical imaging, [17] .

**Semi-supervised Learning:** Semi-supervised learning methods are applied to datasets in which both labelled and unlabeled examples co-exist, with the volume of unlabeled data typically exceeding that of the labeled data. This scenario is common when the cost of labeling data is high which is often the case in the healthcare and medical domains  [18]. Semi-supervised methods use patterns present in the unlabeled data to improve performance over models trained on the labeled data alone. Semi-supervised approaches can also be used iteratively, identifying at each iteration which of the unlabeled points it would be best to acquire the label of by, for example, querying a human expert (Active Learning), [19].

**Reinforcement Learning:** Reinforcement learning (RL) moves away from a static collection of training data to a setup in which an agent learns by receiving feedback as it interacts with its environment, [20].

RL has had high profile success in learning game strategies, where rewards are received at the end of a (variable length) game, [21]. There are various applications of reinforcement learning in healthcare, such as in medical imaging, [22], diagnosis systems, [23] and precision medicine, [24].

**Self-supervised learning**: Self-supervised learning (SSL) has emerged recently, especially in the computer vision field. In this type of learning, deep learning models derive representations of data without requiring labels. This method is particularly useful in the medical field due to the availability of large amount of unlabeled data [25]. Zhao et al [26] proposed the use of SSL method for anomaly detection in medical images.

## Common Machine Learning Models

Machine learning models can be grouped into various categories according to their purpose, algorithm type and how they work and learn, [10]. Table 1 and Table 2 summarizes a set of the most common types of machine learning models categorized by the type of algorithms they use, and how they handle data for both supervised and unsupervised learning.

| Algorithm type | Definition | Models | Examples | Reference |
|---|---|---|---|---|
| Traditional regression / parametric based methods. | Aim at modelling a link between input variables and an output. This enables the prediction of outputs for new examples. Typically, a single parameter needs to be learnt per input feature. | Linear Regression<br><br>Logistic Regression<br><br>Stepwise Regression<br><br>Multivariate Adaptive Regression | Medical images<br><br>landmark<br><br>detection | [27] |
| Instance / non-parametric based classification methods. | These methods use similarity between data instances. Predictions are made by summarizing the outputs for training examples. Weights are increased for data that is more similar to new observation. | K-nearest Neighbor (KNN)<br><br>Self-Organising Map (SOM)<br><br>Support Vector Machines SVM | Decision making in mental health system using SOM | [28] |
| Tree-based Algorithms | Methods that make decisions based on traversing a tree. The value of features at each node is used to decide the direction of the decision. Typically, recursive search methods are used to successively grow trees, or select nodes. | Classification and Decision Tree<br><br>Chi-squared Automatic Interaction Detection (CHAID)<br><br>Conditional Decision Trees | Analysis of adverse drug reaction using CHAID | [29] |
| Bayesian Algorithms | Probabilistic algorithms based upon Bayesian statistical principles. These methods operate by using data to update prior probabilities (of e.g. class membership) to posterior probabilities. | Naïve Bayes<br><br>Bayesian Belief Network (BBN)<br><br>Gaussian Naïve Networks | Modelling fetal mortality | [30] |
| Neural Networks and Deep learning | Neural Networks including Deep learning methods are inspired by biological neural networks. They consist of an input layer, connected to hidden layers, followed by an output later. Each layer consists of a number of neurons. Deep neural networks have architectures with many hidden layers and provide state-of-the-art | Convolution Neural Networks (CNN),<br><br>Recurrent Neural Networks (RNNs) - for example | CNN for the classification of skin cancer<br><br>LSTM for predictive medicine from EHR. | [31]<br><br>[32] |

| | performance for complex data such as images, text and audio. | Long Short-Term Memory networks (LSTM), Auto-Encoders, Deep Belief Network (DBN) | Genomic data imputation using auto encoders | |
|---|---|---|---|---|
| Natural Language Processing (NLP) | NLP aims at developing algorithms to understand and process human languages. Developed method aim to process and understand language in the form of text or speech and is widely used in the medical field. There exist various methods in NLP such as the most classical statistical and rule based approach, machine learning or more recently the use of transformers | Computational linguistics Statistical and machine learning models Deep learning and transformers | Computational linguistics to extract cancer phenotypes Use of BERT in EHR for disease prediction | [33] [34] |

*Table 2 Unsupervised Machine learning Categories*

| Algorithm type | Definition | Models | Examples | Reference |
|---|---|---|---|---|
| Non-parametric Clustering Algorithms | Group the data observations according to similarity between them. This results in a partition of the observations. | Hierarchical Clustering Kernelized K-means | Alzheimer structural imaging phenotype Detection using hierarchical clustering. Diabetes diagnosis using K-means clustering | [35] |
| Parametric Clustering Algorithms | Defines group of data using parametric model. Data is assigned to cluster using the largest prior probability | K-means K-medians Statistical Mixture models Gaussian Mixture Models | Diabetes diagnosis using K-means clustering | [36] |
| Admixture models | Whereas clustering approaches assume that each data point belongs to a single cluster, admixture approaches allow each data point to take contributions from multiple clusters (aka topics). Popular in natural language processing. | Latent Dirichlet Allocation | Modelling tabular healthcare data | [37] |
| Dimensionality reduction | Dimensionality reduction algorithms compress high-dimensional data into a low-dimensional representation. This is achieved by preserving all useful structure of the data in an unsupervised way. The resulting low dimensional representations are used for data exploration and visualization. They can also be used as features for other machine learning models | Principal Component Analysis (PCA)<br><br>Multi-dimensional Scaling (MDS)<br><br>T-Stochastic Neighbour Embedding (TSNE) | Brain tumor segmentation | [38] |
| Neural Networks and Deep Learning | These methods have been defined in Table 1. They can have both supervised or unsupervised learning | | | |

# Machine Learning model development

The development of a machine learning model requires many steps, from data collection through to final model deployment, as depicted in Figure 1. For a model trained within a TRE all steps apart from initial collection and deployment would take place within the TRE, and deployment would necessitate removal of the model from the TRE.

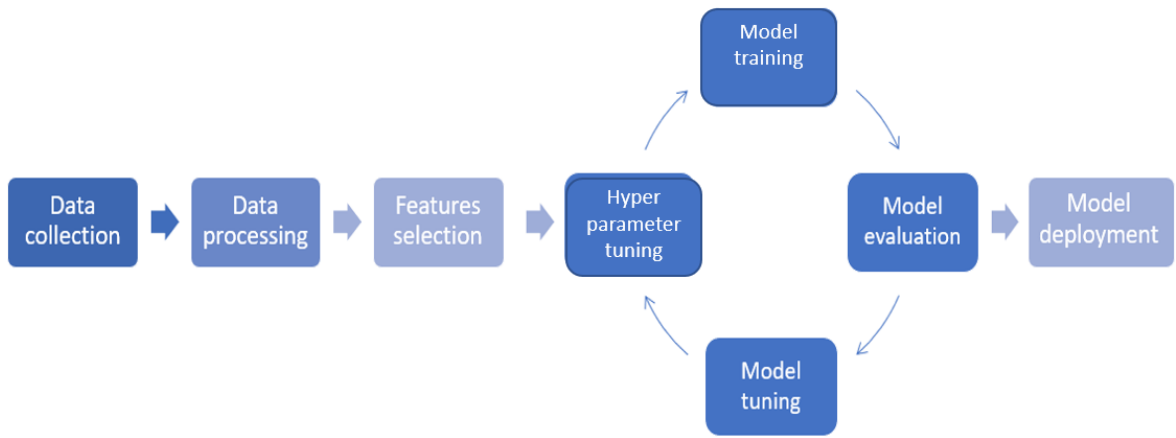

*Figure 1 - Machine Learning Model Development Pipeline*

All ML models typically have parameters that are optimized via training, as well as hyper-parameters that are usually set based on expert or domain knowledge or through trial and error (e.g., setting learning rates for neural network models). Thus, finding the model with the best estimated performance for a desired task is often an iterative process.

A key step in model development, particularly for supervised models, is validation of the model's performance. Model performance should not be evaluated only on training data but also on data unseen during training. Performance on test data is used to estimate how well the model can generalise to unseen data, and hence how well it might perform once deployed. The validation process also highlights when models have simply memorized the training data rather than learning useful structure, an issue known as overfitting, [39].

## Outputs of a Machine Learning training process

A researcher wishing to deploy a ML model trained within a TRE will likely want to export two types of output from the TRE: performance results and analysis, and the trained model. The former would fall within the remit of traditional output checking. However, the latter would not, and we will now explore what kind of information would be held within a trained model file:

### Architecture:

This is most relevant in neural network (and therefore deep learning) models where the architecture defines the number of nodes in each layer, how the nodes are connected, and the activation functions used in each layer. In our definitions we consider architecture to be defined by hyper-parameters, and set *before* training so, e.g., the definitions of the nodes (features and thresholds) in a decision tree are *parameters* rather than *architecture*.

### Parameters:

A model's parameters are those variables within a model that are optimised during training. In the example of a logistic regression, these are the weights that each input variable is multiplied by in the decision function. For a neural network, these are the weights on the connections between neurons.

### Configuration:

Many models save the details of any additional parameters that were set by the user for training, possibly including some information regarding the dataset that was used for training, for example variable names.

### Optimizer and its state:

Models that require optimisation in the training phase will often store the state of the optimiser at the optimum. This can include which optimizer was used, any associated parameters (e.g., learning rate), how many iterations the optimiser ran for, and why it terminated (convergence, the maximum number of iterations was reached, etc). Saving the optimizer state can be useful if further trained if desired.

Subset of training data:

Some algorithms require training examples to make predictions (e.g., K-nearest neighbours and Support Vector Machines). Where this is the case, these training examples would be saved within the model file.

## Vulnerability of trained ML models

There are various stages in the ML development / deployment cycle when the models are vulnerable to attack. For example, the training phase can be compromised if training sets are poisoned and deployed systems can be subject to adversarial attacks. Although these threats are important, we are interested here in the specific risks around disclosure of personal data once a model is outside a TRE, which can comes through membership inference and model inversion attack, [40], [41]. Those attacks represent the most common attacks on ML models. For a comprehensive review of all vulnerabilities of trained ML models, the reader is referred to [1].

Below, we describe membership inference and model inversion attacks in more detail. In both cases, we assume the attack is being undertaken by a malicious actor (the attacker) outside the TRE who has access to the trained model, either directly, or through an interface through which they can query it with data and record its predictions. Note that the attacks we consider require some effort on the part of the attacker (computational or otherwise). This is different to the form of breaches in which sets of record are made available in 'raw form' (the classic 'laptop on a train' scenario and more modern variants) – which should be covered by standard data management policies.

## Membership inference attacks

In a membership inference attack, an attacker has access to some particular data instances and attempts to determine whether they were part of to the data used to train the model. This attack model was first introduced by, [42].

Membership inference attacks leverage the observation that models will often make more confident predictions on data that they were exposed to in training than unseen data. Therefore, high predictive confidence can help to infer the likelihood of a data point belonging to a training set. Overconfidence on training examples is associated with model overfitting, [43] which can be the result of poor model architecture, inappropriate training or too few training examples. This has been demonstrated by [44], who have also experimented on membership inference attacks on federated learning systems where a model is trained on data from various locations simultaneously. To perform membership inference attacks, an attacker will often train an attack model with the predictive probabilities that the model they are attacking (the target model) produces for data that was and was not used for training. As they will not have the data that was actually used for training, they typically train their own versions of the target model (known as shadow models). Although it has been shown that attacks are still possible based upon shadow models which share neither architecture nor training data with the target model [45], the more information the attacker has about the original data and the target model, the more likely they are to be able to succeed.

Membership inference attacks can also be applied to NLP applications such as in [46], where Carlini et al were able to identify training examples from large language models such as GPT-2. They demonstrated that even without overfitting, large language models such as GPT-2 can still memorize sensitive data. Similarly, Vakili et al [47] investigated the privacy preservation of language models such as BERT in clinical data.

## Model inversion attacks

Model inversion attacks, (also known as attribute inference attacks) attempt to infer aspects of the input training data. These attacks can be particularly dangerous for private and confidential data, [48]. For example, an attacker might attempt to infer some particularly sensitive model inputs for one or more individuals based upon known values for the other inputs, and the output prediction, which may be straightforward to obtain.

Various types of inversion attack have been published in the literature. Fredrikson, [49] was the first to introduce the concept of model inversion by showing how a model that predicted drug dosage could be inverted to leak sensitive data about individuals on which it was trained.

Nigesh et al [50] applied model inversion attacks on deep learning model for 3D medical images segmentation tasks. They analyzed the vulnerabilities of models trained on brain images.

Model inversion attacks can benefit from an access to the model type and architecture, where attackers attempt to reproduce the parameters or functionality of a model. For example, if an attacker knows that the model is a logistic regression classifier, then they know the structure and, by presenting sufficient diverse inputs to the model and recording the outputs, can construct a series of equations from which the regression weights can be reverse engineered. If attackers are just interested in mimicking the functionality, they can present many input examples, record the outputs and train a completely new model.

## Challenges in The Disclosure of Machine Learning Models from TREs

Based on the background presented in the previous sections, we believe that there are significant challenges that must be addressed before TREs are able to safely allow trained ML models to be exported. In this section, we summarise what we believe these challenges are, splitting them into three categories:

Users

Threats around trained ML model disclosure can emerge from different users and actors, of whom we identify four.

**Non-malicious researchers:** It is likely that many TRE users will have little awareness of possible threats faced by trained ML models. For example, it is easy to imagine a researcher training a Support Vector Machine (SVM), unaware that the saved SVM has to include a copy of at least part of the training dataset for it to operate, rendering it prone to membership inference attacks and therefore a considerable disclosure risk. Similarly, although most ML developers would understand the concept of over-fitting, they would likely not realise the link between over-fitting and vulnerability to membership inference attack.

**Malicious users:** these are users who deliberately hide data inside disclosed models and outputs. They directly cause a data breach, by dissimulating data inside other outputs. To a certain extent, malevolent behavior is guarded against through the existing TRE safeguarding procedures, [2]. However, these guidelines were designed for aggregated results (tables, plots, or summary statistics), in which the possibility of hiding large quantities of data was insignificant. This changes with the disclosure of ML models, where files being disclosed could be large and not necessarily human readable.

**External attackers:** The third actor is an external attacker who has access to trained models after they have been disclosed from the TRE, through either model deployment (e.g. via an accessible application programming interface; API) or model sharing. Attacks can be carried out by these actors in various forms as described in the previous section.

**TRE output checkers**: The final actors are the TRE output checkers themselves. Given the relative novelty of ML approaches, it is highly likely that the majority of TRE output checking staff have little or no familiarity with these approaches and are thus being asked to check things that they do not understand, to assess risks that they also do not understand.

## Data

The application of machine learning in healthcare represents more challenges when disclosing models from TRESs. Healthcare data is very heterogenous and unstructured ranging from Electronic Health Records (EHR), medical images to large genomic databases. Within TREs, data is safe-guarded and follows strict protocols such as de-identification [51], anonymization, pseudonymization [52]. Anonymization aims at removing any personal identifiers from data resulting to data being completely unidentifiable. De-identification consists of removing any personal direct identifiers without fully preventing any information link that could lead to direct or indirect identification of individuals. Pseudonymization on the other hand represents the process of replacing certain identifiers, where data can no longer be linked to personal identifiers without the recourse to further processing [53] .

Depending on the data types being considered, TRE operators need to adopt appropriate methods for removing personal data. Whereas in a traditional analysis, an output checker could see the output and assess if the process followed by the researcher was exposing sensitive data, this is substantially more challenging with a trained ML model.

## ML models

As described in the previous sections ML models are vulnerable to attacks that can potentially lead to data privacy breaches. Such attacks are possible even when the attacker has only *black box* access to the model, where they cannot access the model file itself but only an interface or API. Attackers can therefore *use* the model (present inputs and be provided with outputs) but cannot observe the inner workings of the model. In some cases, an attacker might get access to the model file (a *white box* attack) and therefore get access to model parameters and architecture. White box attackers can therefore both use and inspect the model, [39]. In general, white box access confers greater risk. However, black box attacks can also represent a significant risk but require more effort to be successful.

The challenge of disclosing ML models varies between different models and training regimes. Some models are more prone to attacks than others although model configuration (architecture and setting of model hyper-parameters) also plays a significant role. Disclosing an SVM or a tree-based algorithm may be riskier than disclosing more complex models, [54]. More work is required to better understand the risk of a wide range of models and configurations.

## Legal and Regulatory Issues

The personal data used in TREs, and which may be exported in a trained ML model, will be governed by data protection law if it exists in the particular jurisdiction. Internationally, the European Union's General Data Protection Regulation (GDPR) is the most prominent data protection framework, and the UK implemented it into domestic law before it left the EU as a Member State in the Data Protection Act 2018. Given our focus on UK-based TREs, we proceed with some key points about its application.

The key challenge is responsibility for a possible data breach resulting from export of a ML model, where the output checker does not have access to human readable outputs.

Understanding that the disclosure of a trained ML model, which works differently to the disclosure of standard statistics (e.g. graphs and tables etc.), is important.  The security of processing, is a responsibility attributed to the controller and the processor, as outlined in Article 32 of the GDPR.  The controller and processor must take "state of the art", technical and organisational measures into account.  These include "appropriate" measures commensurate with the risks, in terms of (a) pseudonymisation and encryption, (b) ongoing resilient systems and services, (c) ability to restore personal data (PD), and (d) regular testing for effectiveness of the security of the processing [55].

For a trained ML model, the chain of responsibility may be unclear where the initial data controller and TRE processor are no longer involved when the ML model has been transferred out of the TRE, and if the responsibility for any potential future personal data breach has not been identified nor allocated correctly in the terms of agreement that a TRE operator has made with a TRE user. Assessing the level of risk as to whether the trained ML model and/ or its output are inextricably linked to potentially disclosive personal data it was trained on will determine how the law categorises the model. We consider there are 3 categories:

1. A trained ML model can be considered to only contain anonymous data and therefore GDPR does not apply [56].

2. A trained ML model is considered to potentially contain pseudonymised personal data, therefore requiring specific technical and organisational measures to ensure the processing is GDPR compliant. In particular, whether appropriate data security measures have been adopted to reflect the risk of data breach.

3. A trained ML model is considered to carry more risk of including  disclosive personal data (pseudonymised) therefore requiring extra layers of protection in terms of the transfer of responsibility, obligations and rights to a new data controller / processor, with prior written authorisation of the controller, or indeed the retention of responsibility, obligations and rights by the original data controller in the contract [57].

## Risk analysis in disclosing models from TREs

In the previous sections, we have given an overview of machine learning models, their different types, applications in healthcare and their potential vulnerabilities. We have also described challenges faced when disclosing machine learning models from TREs from different aspects such as data types, user types, and the particular machine learning model(s) being used.  In this section, we describe we describes different factors that may affect the risk levels in disclosing trained ML Models.

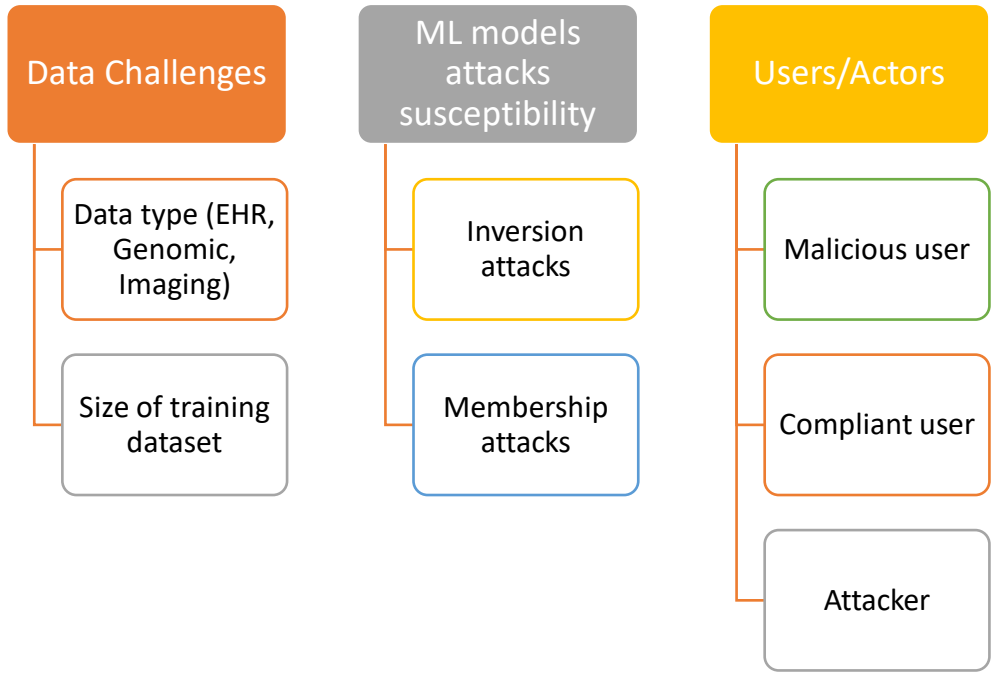

*Figure 2 - Machine Learning Challenges Types*

It is clear that preserving data privacy when disclosing ML models is challenging. For a model to be effective, it must retain some aspects of the data on which it was trained, [41].

Disclosing ML models from safe environments therefore carries risks that must be accounted for analysed and mitigated. We have identified various factors that may affect the level of risks such as data types, machine learning type or the attack type. Figure 2 summarizes the above-mentioned factors.

- ***Data types***: The heterogeneity of healthcare and medical data contribute to the different levels of risks associated with disclosing models from TREs. In fact, some types of data can pose greater risks than others. Clinical text, and clinical images have different level of risks in re-identification of personal and private data, for example an X-ray of a knee is less disclosive than clinical phenotype for rare diseases.

- ***Machine learning types***: In addition to different data types. Using different machine learning models can alter the risk level. Some models pose more risks than others as demonstrated in [46], where large language models may remember data even if overfitting is reduced.

- ***Actor types:*** Different actors will have different effects on preserving privacy in machine learning models. Outsider attackers will aim at reconstructing part or all of the training data. This would negatively affect the risk of disclosure of models. Similarly, a user that follows processes to create privacy preserving models would pose lower risk than a less knowledgeable user.

In addition, the combination of different factors contributes to the level of risk. For example, disclosing an SVM model trained on brain MRI data represents a higher risk of re-identification than disclosing a deep learning model using genomic data. Medical images such as brain images can contain hidden biometrics, [58], and pose a higher risk to privacy if reconstructed via a membership inference attack. [59].

To assess the risk of disclosure, TREs need to consider the severity of the different factors (alone or in combination) and the impact of a data breach, which would primarily concern re-identification of individuals and subsequent loss of trust in TREs. This can be achieved by conducting thorough risk analysis before the disclosure of machine learning models.

Future of disclosure control of machine learning models in TREs

In this paper, we have attempted to draw attention to a new problem facing TREs: the output of trained ML models. Our previous research has demonstrated that TREs are ill-equipped for this task, although it is likely that requests to remove trained models from TREs will become increasingly common. As well as describing TREs, we have provided a brief introduction to Machine Learning, as well as the privacy risks that a trained ML model can present.

We believe that research is urgently needed to form the basis of new output checking procedures for TREs. Detailed risk analysis needs to be conducted to quantify risks using the different factors and challenges outlined in this paper. Various risk model frameworks have been investigated in the literature in domains such as financial, privacy or security, [60]. Model frameworks such as FAIR (Factor Analysis of Information Risk), [61] could be investigated and applied to conduct a risk assessment of ML models disclosure from TREs. FAIR is a risk management model that divides different aspects of risks into different factors that quantify risk into different probabilities. However, this is not a strictly statistical discussion. The nature of TREs means that multiple measures of control are potentially available, and so a high baseline risk may still be acceptable. This is particularly relevant when considering the motivation of attackers.

The challenges and risks described in this paper create new opportunities for interdisciplinary research. Investment in the following open areas of research will play a role in ensuring that TREs can facilitate safe disclosure of ML models:

- ***Data privacy research:*** Various opportunities exist in providing more secure processes at the data level. This could include encryption of the data and enabling the application of machine learning algorithm directly on encrypted data. Methods such as homomorphic encryption allow this. Adding digital watermarks enables data tracking that could help detect privacy leaks [62]. ML models could also be trained on synthetic data. However, this method can only be applied in very limited types of data and applications applying synthetic data in the medical field can lead to algorithm bias, [63]. TREs can also provide guidelines for researchers and users to ensure data privacy throughout the whole ML models life cycle; where users need to insure data privacy during training, input privacy, output privacy and models privacy.

- *ML Model evaluation*: The community needs large scale evaluation studies to attempt to uncover the vulnerabilities of common ML models. Although there has been progress in this area, much more experimental work is needed to assess the wide range of models that researchers may wish to use in TREs. As ML researchers develop new methods of training models that aim to improve privacy (e.g., differentially private model training [64] , [65], federated learning [48]) these must also be evaluated. The results of all of these analyses must be made accessible to TREs controllers.

- *Automatic ML model Privacy assessment* Quantifying risks from machine learning models is impossible using human controls only.  Providing TREs and researchers with tools that helps assess risk factors for models can help quantify, manage and mitigate risks in disclosure of such models. Various machine learning attack metrics tools have been proposed in the literature such as those in, [14].

- **AI responsibility and accountability**   Exploring areas such as accountability, explainability [66], and fairness from both a technical and legal point of view.  In addition to explainable Artificial Intelligence (XAI), the legal community will need to be able to prove who and / or what was responsible for a data breach (causality) [66].  If we consider that the GDPR allocates responsibility to a person (controller / processor / sub-processor) when determining who is responsible for a breach in testing the effectiveness of the de-identification of personal data, it is clear that current UK GDPR law regulates for responsibility to be attributed to a person [67].

- **Model based mitigation strategies and guidelines:** Develop methodologies for mitigating risks linked to model disclosure such as restricting types of models that can be used. Some model types such as KNN and SVM represent a higher risk and need a tighter control.

- **Effective manual disclosure procedures**: This could include model / code inspections prior to disclosure. This requires experts within the disclosure teams, and the technical infrastructure to generate and store code snapshots for inspection. Ideally, the disclosure team would also use the model, independently evaluating model performance on both the data that was provided to the researcher and a subset of data that they had never used. For example, in a large dataset, 10% of the examples could be withheld completely from the researcher for use by the disclosure team.

# Funding

This project was in part supported by MRC and EPSRC program grant: InterdisciPlInary Collaboration for efficient and effective Use of clinical images in big data health care Research: PICTURES [grant number MR/S010351/1].

This work was in part supported by Health Data Research UK (HDR UK: 636000/ RA4624) which receives its funding from HDR UK Ltd (HDR-5012) funded by the UK Medical Research Council, Engineering and Physical Sciences Research Council, Economic and Social Research Council, Department of Health and Social Care (England), Chief Scientist Office of the Scottish Government Health and Social Care

Directorates, Health and Social Care Research and Development Division (Welsh Government), Public Health Agency (Northern Ireland), British Heart Foundation (BHF) and the Wellcome Trust.

This work was in part supported by the Industrial Centre for AI Research in digital Diagnostics (iCAIRD) which is funded by Innovate UK on behalf of UK Research and Innovation (UKRI) [project number: 104690].

This work was funded by UK Research and Innovation Grant Number MC_PC_21033 as part of Phase 1 of DARE UK (Data and Analytics Research Environments UK) programme, delivered in partnership with HDR UK and ADRUK.  The specific project was Guidelines and Resources for AI Model Access from TrusTEd Research environments (GRAIMatter).